\documentclass[prd,eqsecnum,twocolumn,amsfonts,amssymb]{revtex4}

\usepackage{graphicx}

\usepackage{bm}
\usepackage{framed}

\newcommand{\beq}{\begin{equation}}
\newcommand{\eeq}{\end{equation}}
\newcommand{\beqs}{\begin{eqnarray}}
\newcommand{\eeqs}{\end{eqnarray}}

\newcommand{\drawsquare}[2]{\hbox{%
\rule{#2pt}{#1pt}\hskip-#2pt%  left vertical
\rule{#1pt}{#2pt}\hskip-#1pt%  lower horizontal                       
\rule[#1pt]{#1pt}{#2pt}}\rule[#1pt]{#2pt}{#2pt}\hskip-#2pt%  upper horizontal  
\rule{#2pt}{#1pt}}% right vertical
\newcommand{\fund}{\raisebox{-.5pt}{\drawsquare{6.5}{0.4}}}%  fund
\begin{document}

\title{A Comparative Study of Criticality Conditions for
  Anomalous Dimensions using Exact Results in an ${\cal N}=1$
  Supersymmetric Gauge Theory}

\author{Thomas A. Ryttov$^a$ and Robert Shrock$^b$}

\affiliation{(a) \ CP$^3$-Origins, University of Southern Denmark, \\
Campusvej 55, Odense, Denmark}

\affiliation{(b) \ C. N. Yang Institute for Theoretical Physics and 
Department of Physics and Astronomy, \\
Stony Brook University, Stony Brook, NY 11794, USA }

\begin{abstract}

  Two of the conditions that have been suggested to determine the
  lower boundary of the conformal window in asymptotically free gauge
  theories are the linear condition, $\gamma_{\bar\psi\psi,IR}=1$, and
  the quadratic condition,
  $\gamma_{\bar\psi\psi,IR}(2-\gamma_{\bar\psi\psi,IR})=1$, where
  $\gamma_{\bar\psi\psi,IR}$ is the anomalous dimension of the
  operator $\bar\psi\psi$ at an infrared fixed point in a theory. We
  compare these conditions as applied to an ${\cal N}=1$
  supersymmetric gauge theory with gauge group $G$ and $N_f$ pairs of
  massless chiral superfields $\Phi$ and $\tilde \Phi$ transforming
  according to the respective representations ${\cal R}$ and $\bar
  {\cal R}$ of $G$.  We use the fact that $\gamma_{\bar\psi\psi,IR}$
  and the value $N_f = N_{f,cr}$ at the lower boundary of the
  conformal window are both known exactly for this theory. In contrast
  to the case with a non-supersymmetric gauge theory, here we find
  that in higher-order calculations, the linear condition provides a
  more accurate determination of $N_{f,cr}$ than the quadratic
  condition when both are calculated to the same finite order of
  truncation in a scheme-independent expansion.
    
\end{abstract}

\maketitle

% =======================================================================

% section 1
\section{Introduction}
\label{intro_section}

There has been considerable interest in asymptotically free gauge
theories that have matter content such that they exhibit
renormalization-group flows from the deep ultraviolet (UV) to infrared
(IR) fixed points (IRFPs) \cite{b2,bz}.  At the infrared fixed point,
the beta function vanishes, so the theory is scale-invariant, and is
inferred to be conformally invariant \cite{scalecon}, whence the term
``conformal window''.  With no loss of generality, one may restrict to
massless matter fields, since if a matter field had a nonzero mass
$m_0$, one would integrate it out of the effective low-energy theory
that is relevant for momentum scales below $m_0$ in the flow to the
infrared limit.  The properties of a theory at an infrared fixed point
in this conformal window are of fundamental interest.  Among these are
the scaling dimensions $D_{\cal O}$ of various (gauge-invariant) local
operators, ${\cal O}$, such as $\bar\psi\psi$ and ${\rm
  Tr}(F_{\mu\nu}F^{\mu\nu})$, where $\psi$ and $F_{\mu\nu}$ denote
fermion and gauge field-strength operators. Owing to the gauge
interactions, the scaling dimension of an operator ${\cal O}$ differs
from its free-field value, $D_{{\cal O},free}$: $D_{\cal O} = D_{{\cal
    O},free} - \gamma_{\cal O}$, where $\gamma_{\cal O}$ is the
anomalous dimension of ${\cal O}$. Higher-loop calculations of
anomalous dimensions at an IR fixed point in the conformal window have
been performed in a number of works, including
\cite{bvh}-\cite{dexm2}, using both conventional series expansions in
powers of the gauge coupling at the IR fixed point and in powers of a
scheme-independent expansion variable. Inputs for
renormalization-group functions utilized in this work included those
in \cite{b5}-\cite{c5}. Extensive measurements of anomalous dimensions
have been carried out using lattice simulations; some of these works
are \cite{afn}-\cite{simons}.

As one decreases the matter content, the value of the gauge coupling
at the IRFP, $\alpha_{IR}$, increases, and eventually the theory
changes qualitatively with the disappearance of this conformal IR
fixed point.  A commonly studied example is a non-Abelian gauge theory
(in $d=4$ spacetime dimensions at zero temperature) with gauge group
$G$ and $N_f$ copies (``flavors'') of massless Dirac fermions
transforming according to a representation $R$ of $G$.  One arranges
that $N_f$ is smaller than an upper ($u$) bound, $N_{f,u}$, depending
on $G$ and $R$, so that the theory is asymptotically free. As $N_f$
decreases below $N_{f,u}$, the theory exhibits the aforementioned
conformal IRFP, and the lower boundary of the conformal window occurs
as $N_f$ decreases through a critical value denoted $N_{f,cr}$
\cite{atw}. Generalizations of this with several fermions transforming
according to different representations have also been studied
\cite{dexm,khl,jwlee,dexm2,fx422,deldebbio,fx444}, but here it will be
sufficient for our analysis to restrict our consideration to the case
of matter fields transforming according to a single representation of
the gauge group.

In addition to its importance in the context of formal quantum field
theory, a determination of $N_{f,cr}$ is important for the analysis of
gauge theories with $N_f$ slightly less than $N_{f,cr}$, since in
choosing such a theory to study, one needs to know at least the
approximate value of $N_{f,cr}$. A theory with $N_f$ slightly below
$N_{f,cr}$ has a gauge coupling that runs slowly over a large range of
momentum scales, due to an approximate IR zero in the beta function,
but eventually becomes large enough to produce spontaneous chiral
symmetry breaking and associated dynamical breaking of the approximate
dilatation invariance. As a result, these theories (often called
``walking'' or quasi-conformal theories) feature an approximate Nambu-Goldstone
boson, the dilaton, as has been confirmed by lattice simulations
\cite{latkmi_2017,lsd_2019,kuti_dilaton}.  Since the mass of a
Nambu-Goldstone boson is protected against large radiative
corrections, models incorporating this physics thus have the potential to
address the Higgs mass hierarchy problem \cite{dilaton}.

Two of the conditions that have been suggested to determine the lower
boundary of the conformal window in asymptotically free gauge theories
are the linear critical condition ($\gamma$CC),
$\gamma_{\bar\psi\psi,IR}=1$, and the quadratic critical condition,
$\gamma_{\bar\psi\psi,IR}(2-\gamma_{\bar\psi\psi,IR})=1$
\cite{alm}-\cite{zwicky}.  As is evident from the fact that the
quadratic critical condition can be rewritten equivalently as
$(\gamma_{\bar\psi\psi,IR}-1)^2=0$, it has a double root at
$\gamma_{\bar\psi\psi,IR}=1$ and hence is formally identical to the
linear $\gamma$CC.  However, these two critical conditions yield
different predictions for $N_{f,cr}$ when using, as input, a
finite-order series expansion for $\gamma_{\bar\psi\psi,IR}$. In
non-supersymmetric gauge theories, the quadratic condition has been
found to converge faster as a function of the order to which this
series for $\gamma_{\bar\psi\psi,IR}$ is computed \cite{khl,jwlee}. An
interesting question concerns how general this difference is; i.e., is
it the case that the quadratic critical condition will also yield more rapid
convergence than the linear critical condition in other theories?

In this paper we investigate this question, using as our theoretical
laboratory an ${\cal N}=1$ supersymmetric gauge theory with gauge
group $G$ and $N_f$ pairs of massless chiral superfields $\Phi$ and
$\tilde \Phi$ transforming according to the respective representations
${\cal R}$ and $\bar {\cal R}$ of $G$.  We take advantage of the key
fact that for this theory one has exact results for
$\gamma_{\bar\psi\psi,IR}$ and $N_{f,cr}$
\cite{nsvz,seiberg,shifman_review,terning}.

This paper is organized as follows.  In Sect. \ref{framework_section}
we review some relevant background concerning the
${\cal N}=1$ supersymmetric gauge theory and our calculational methods.
Section \ref{gamma_crit_section} contains a discussion of the linear
and quadratic critical conditions on $\gamma_{\bar\psi\psi,IR}$. 
Our calculational results on the comparison of these conditions for
the supersymmetric theory are presented in Sect. \ref{results_section}.
Our conclusions are summarized in Sect. \ref{conclusions_section}. 

% ===================================================================

\section{Background on the ${\cal N}=1$ Supersymmetric Gauge Theory and
Calculational Methods}
\label{framework_section}

In this section we briefly review some relevant background and our
calculational methods.  We consider a vectorial ${\cal N}=1$
supersymmetric gauge theory (in $d=4$ spacetime dimensions) with gauge
group $G$ and matter content consisting of $N_f$ flavors of massless
chiral superfields in the fundamental and conjugate fundamental
representations, denoted as $\Phi$ and $\tilde \Phi$ (with color and
flavor labels implicit here).  In terms of component fields, the
chiral superfield $\Phi$ has the decomposition
\beq
\Phi = \phi + \sqrt{2} \, \psi \theta + F \theta\theta \ , 
\label{components}
\eeq
where $\psi$ is taken as a left-handed Weyl fermion, $\theta$ is
an anticommuting Grassmann variable, and $F$ is a non-dynamical
auxiliary field.

We denote the running gauge coupling as $g=g(\mu)$, where $\mu$ is the
Euclidean energy/momentum scale at which this coupling is measured,
and define $\alpha(\mu) = g(\mu)^2/(4\pi)$.  As noted above, we
restrict consideration of this theory to the range of $N_f$ where it
is asymptotically free. Owing to this, its properties can be computed
perturbatively in the UV limit at large $\mu$, where $\alpha(\mu) \to
0$.  The dependence of $\alpha(\mu)$ on $\mu$ is described by the
renormalization-group (RG) beta function,
\beq
\beta = \frac{d\alpha(\mu)}{d\ln\mu} \ .
\label{beta}
\eeq
The argument $\mu$ will generally be suppressed
in the notation.  The series expansion of $\beta$ in powers of
$\alpha$ is
\beq
\beta = -2\alpha \sum_{\ell=1}^\infty b_\ell \, a^\ell \ ,
\label{beta_series}
\eeq
where
\beq
a \equiv \frac{g^2}{16\pi^2} = \frac{\alpha}{4\pi} \ , 
\label{a}
\eeq
and $b_\ell$ is the $\ell$-loop coefficient.  We restrict here to
mass-independent, supersymmetry-preserving
regularization/renormalization schemes and to gauge-independent scheme
transformations. The first two
coefficients in (\ref{beta_series}) are \cite{jones75}
\beq
b_1 = 3C_A - 2T_f N_f
\label{b1}
\eeq
and \cite{machacek}-\cite{jones84}

\beq
b_2=6C_A^2-4(C_A+2C_f)T_fN_f \ ,
\label{b2}
\eeq
where $C_A$, $T_f$, and $C_f$ are group invariants
\cite{group_invariants}.  These coefficients $b_1$ and $b_2$ are
scheme-independent, while the $b_\ell$ with $\ell \ge 3$ are
scheme-dependent.  With an overall minus sign extracted, as in
Eq. (\ref{beta_series}), the condition of asymptotic freedom is that $b_1 >
0$, and thus $N_f< N_{f,u}$, where the upper bound on $N_f$ is
\beq
N_{f,u} = \frac{3C_A}{2T_f} \ .
\label{Nu}
\eeq
Note that if $N_f=N_{f,u}$ so that $b_1=0$, then the two-loop
coefficient has the negative value $b_2=-12C_fC_A$, so (with the minus
sign prefactor in Eq. (\ref{beta_series})) the theory is not
asymptotically free.  This is the reason that we require the strict
inequality $N_f < N_{f,u}$ for asymptotic freedom rather than the
condition $N_f \le N_{f,u}$.

A number of additional exact results have been established about the
IR phase structure of the theory
\cite{nsvz,seiberg,shifman_review,terning}. We briefly summarize some
relevant properties here. For a general gauge group $G$ and
representation ${\cal R}$, if $N_f$ is in the conformal-window (CW) 
interval
\beq
    {\rm CW}: \quad N_{f,cr} \le N_f < N_{f,u} \ , \quad i.e., \quad  
\frac{N_{f,u}}{2} \le N_f < N_{f,u} \ , 
\label{cw}
\eeq
where 
\beq
N_{f,cr}= \frac{3C_A}{4T_f} = \frac{N_{f,u}}{2} \ ,
\label{Nell}
\eeq
the theory flows from the UV to an IR fixed point of the
renormalization group. (The CW interval is also commonly called the 
non-Abelian Coulomb phase.)  

In general, the expressions in Eqs. (\ref{Nu}) and (\ref{Nell}) for
$N_{f,u}$ and $N_{f,cr}$ are not necessarily integers.  In cases where
$N_{f,u}$
or $N_{f,cr}$ is not an integer, one implicitly treats it as a formal
result applicable in the framework in which one generalizes $N_f$ from
the non-negative integers to the non-negative real numbers.  This will
not be important for our present analysis, which focuses on a 
comparison of the relative accuracies of linear and quadratic $\gamma$
critical conditions when used with finite-order perturbative
anomalous-dimension inputs.  However, for reference, we give some illustrative
examples for the case $G={\rm SU}(N_c)$. If ${\cal R}=F$, the
fundamental representation, then $N_{f,cr}=(3/2)N_c$, which is integral
if and only if $N_c$ is even. If ${\cal R}=Adj$, the adjoint representation,
then $N_{f,u}=3/2$ and $N_{f,cr}=3/4$. Finally, if ${\cal R}$ is the rank-2
symmetric or antisymmetric tensor representation (denoted $S_2$ and $A_2$,
respectively), then ,
$N_{f,u}=2N_{f,cr}=3N_c/(N_c \pm 1)$, where the upper (lower) sign applies for
$S_2$ and $A_2$.

With $b_1 > 0$ for asymptotic freedom, the condition that this
two-loop beta function should have an IR zero is that $b_2 < 0$, which is
that $N_f > N_{f,b2z}$, where 
\beq
N_{f,b2z} > \frac{3C_A^2}{2T_f(C_A+2C_f)} \ .
\label{b2negative}
\eeq
As we discussed in \cite{bfs} (see also \cite{susy_inspired,bfss}),
$N_{f,b2z}$ may be larger or smaller than $N_{f,cr}$, depending on the
chiral superfield representation ${\cal R}$

For a general gauge group $G$, the ${\cal N}=1$ theory under
consideration here, with $N_f$ flavors of chiral superfields $\Phi$
and $\tilde \Phi$ in the representations ${\cal R}$ snd $\bar {\cal
  R}$, respectively, is invariant under a classical continuous global
($cgb$) symmetry
\beqs
G_{cgb} &=& {\rm U}(N_f) \otimes {\rm U}(N_f) \otimes {\rm U}(1)_R \cr\cr
       &=& {\rm SU}(N_f) \otimes {\rm SU}(N_f) \otimes
{\rm U}(1)_V \otimes {\rm U}(1)_A \otimes {\rm U}(1)_R \ , \cr\cr
&&
\label{gcl}
\eeqs
where the first and second U($N_f$) groups consist of operators acting on
$\Phi^j$ and $\tilde\Phi_i$, respectively, with $i,j=1,...,N_f$, and the
$R$-symmetry group U(1)$_R$ is defined by the following commutation
relations
\beq
[Q_\alpha,R]=Q_\alpha \ , \quad [Q_\alpha^\dagger,R]= -Q_\alpha^\dagger \ ,
\label{qrrel}
\eeq
where the $Q_\alpha$ and $Q_\alpha^\dagger$ are the
generators of the supersymmetry transformations (with $\alpha$ 
spinor index here). The U(1)$_A$ symmetry is anomalous, due to
instantons, so the actual non-anomalous continuous global symmetry of the
theory is
\beqs
G_{gb} = {\rm SU}(N_f) \otimes {\rm SU}(N_f) \otimes {\rm U}(1)_V \otimes
{\rm U}(1)_R \ .
\label{gbl}
\eeqs
This symmetry is exact at an IR fixed point in the conformal window. 
The representations of the matter chiral superfields under the gauge and global
symmetry groups are listed in Table \ref{sgt_table} for the generic case in
which the representation ${\cal R}$ is complex. 

% ===============================================

\begin{table}
  \caption{\footnotesize{
Matter content of a vectorial $\mathcal{N}=1$ supersymmetric gauge theory
with gauge group $G$ and
matter content consisting of $N_f$ massless chiral superfields $\Phi$
and $\tilde \Phi$ transforming according to the representations
${\cal R}$ and $\bar {\cal R}$, respectively. The symmetry groups
  correspond to those in Eq. (\ref{gbl}). }}
\begin{center}
  \begin{tabular}{|c||c|c|c|c|c|}
  \hline
& SU($N_c$) & SU($N_f$) & SU($N_f$) & U(1)$_V$ & U(1)$_R$   \\
 \hline
 $\Phi $   & ${\cal R}$ & $\fund$   &  1  & 1 & $1-[C_A/(2T_fN_f)]$ \\
 \hline
  $\tilde{\Phi}$  &  $\overline{\cal R}$ & 1 &  $\overline{\fund}$ & $-1$ & \
 $1-[C_A/(2T_f N_f)]$ \\
    \hline
  \end{tabular}
  \label{sgt_table}
\end{center}
\end{table}

% ==================================================

We will focus on the gauge-invariant quadratic operator products
of the ``meson'' type, 
\beq
M_i^j= \tilde \Phi_i \Phi^j \ ,
\label{meson}
\eeq
where, as above, $i$ and $j$ are flavor indices and the group indices
are implicit, with it being understood that they are contracted in
such a way as to yield a singlet under the gauge group $G$.  As a
holomorphic product of chiral superfields, $M_i^j$ is again a chiral
superfield.  The bilinear fermion operator product in $M_i^j$ is
$\tilde \psi_i \psi^j \equiv \tilde \psi_{i,L}^T C \psi^j_L$, where
$C$ is the conjugation Dirac matrix, and we use the convention of
writing $\tilde \psi_{i,L}$ and $\psi^j_L$ as left-handed Weyl
fermions.  Because the global symmetry (\ref{gbl}) is exact in the
conformal window, the meson-type quadratic chiral superfields
transform according to (irreducible) representations of the group
$G_{gb}$. The anomalous dimension
of this operator is independent of the flavor indices $i$ and $j$
\cite{gracey}, so in \cite{dexss} and here, we denote its value at the
superconformal IRFP simply as $\gamma_{M,IR}$. Using the fact that 
$\tilde \psi_{i,L}=(\psi^i_R)^c$, the fermion
bilinear in $\tilde \Phi_i \Phi^i$ can be rewritten in the standard form
$\bar\psi_i \psi^i$ of a mass term in a non-supersymmetric theory.
Denoting $\gamma_{\bar\psi\psi,IR}$ as the anomalous dimension of the latter
bilinear, it follows that
\beq
\gamma_{M,IR} = \gamma_{\bar\psi\psi,IR} \ .
\label{gamgam}
\eeq

A closed-form expression for the beta function of this theory was derived
by Novikov, Shifman, Vainshtein, and Zakharov (NSVZ) \cite{nsvz}:
\beq
\beta_{NSVZ} = -\frac{\alpha^2}{2\pi} \bigg [ \frac{b_1 -
  2N_fT_f \, \gamma_M}
{1-\frac{C_A\alpha}{2\pi}} \bigg ] \ .
\label{beta_nsvz}
\eeq

It is convenient to introduce the notation
\beq
x \equiv \frac{N_f}{N_{f,u}} 
\label{xdef}
\eeq
and
\beq
x_{cr} \equiv \frac{N_{f,cr}}{N_{f,u}} = \frac{1}{2} \ .
\label{xcr}
\eeq
Thus, the conformal window is the interval
\beq
\frac{1}{2} \le x < 1 \ .
\label{xcw}
\eeq
One can express the anomalous dimension of an operator such as a
fermion bilinear $\bar \psi \psi$ in a gauge theory as a
series expansion in the squared gauge coupling,
\beq
\gamma_{\bar \psi \psi} = \sum_{\ell=1}^\infty c_\ell \, a^\ell \ ,
\label{gamma_series}
\eeq
where $c_\ell$ is the $\ell$-loop coefficient. As noted above, the
value of this anomalous dimension at an IRFP is written as
$\gamma_{\bar \psi \psi,IR}$.  The one-loop coefficient $c_1$ is
scheme-independent, while the $c_\ell$ with $\ell \ge 2$ are
scheme-dependent.

Physical quantities such as anomalous dimensions at an IRFP clearly
must be scheme-independent. In conventional computations of these
quantities, one first writes them as series expansions in powers of
the coupling, as in (\ref{gamma_series}), and then evaluates these
series expansions with $\alpha$ set equal to $\alpha_{IR}$, calculated
to a given loop order.  However, a (finite-order) series expansion of
this type is scheme-dependent beyond the leading terms.  Scheme
dependence is also present in higher-order perturbative calculations
in quantum chromodynamics (QCD), and its effects have been routinely
addressed in studies comparing perturbative QCD predictions with
experimental data. Formally speaking, these studies were on scheme
dependence in the vicinity of the UV fixed point at zero coupling in
QCD.  Studies of scheme dependence in the different context of an IR
fixed point located away from zero coupling have been carried out in
\cite{bfss}, \cite{sch}-\cite{schemegauge}.  For perturbative series
calculations of anomalous dimensions, it is desirable to use a
formalism in which results calculated to each order are
scheme-independent.

Since $\alpha_{IR} \to 0$ as $b_1 \to 0$ at the upper end of the
conformal window, as it follows that one can reexpress the series
expansion for $\gamma_{\bar\psi\psi,IR}$ in terms of a variable that
is proportional to $b_1$, namely the scheme-independent variable
\cite{bz,gkgg}
\beq
\Delta_f = N_{f,u}-N_f \ . 
\label{Delta}
\eeq
In the present theory,
\beq
\Delta_f = \frac{b_1}{2T_f} \ .
\label{Delta_b1}
\eeq
Scheme-independent calculations of anomalous dimensions of various
operators at an IRFP were carried out in \cite{gtr}-\cite{baryon} for
non-supersymmetric gauge theories, and results were compared with
measured values from lattice simulations. In \cite{dexss} we carried
out corresponding scheme-independent calculations of anomalous
dimensions of several composite superfield operator
products in the present ${\cal N}=1$
supersymmetric theory. In general, the scheme-independent series
expansion for a (gauge-invariant) operator ${\cal O}$ at an IRFP in
the conformal window can be written as
\beq
\gamma_{{\cal O},IR} = \sum_{j=1}^\infty \kappa_{{\cal O},j} \, \Delta_f^j \ .
\label{gamma_IR_Deltaseries}
\eeq
The truncation of this series to $O(\Delta_f^p)$ inclusive is denoted
$\gamma_{{\cal O},IR,\Delta_f^p}$:
\beq
\gamma_{{\cal O},IR,\Delta_f^p} = \sum_{j=1}^p \kappa_{{\cal O},j} \,
\Delta_f^j \ .
\label{gamma_dpdef}
\eeq
Thus, for the operator $M$ we write
\beq
\gamma_{M,IR,\Delta_f^p} = \gamma_{\bar\psi\psi,IR,\Delta_f^p} =
\sum_{j=1}^p \kappa_{M,j} \, \Delta_f^j \ 
\label{gamma_M_dpdef}
\eeq
with
\beq
\kappa_{M,j} = \kappa_{\bar\psi\psi.j}
\label{kapkap}
\eeq

It is convenient to define the reduced scheme-independent expansion
variable 
\beq
y \equiv \frac{\Delta_f}{N_{f,u}} = 1 - \frac{N_f}{N_{f,u}} = 1-x \ . 
\label{Deltabar}
\eeq
Since $1/2 \le x < 1$ in the conformal window (cf. Eq. (\ref{xcw})), it
follows that in the conformal window $y$ takes values in the
range 
\beq
{\rm CW}: \quad 0 < y \le \frac{1}{2} \ ,
\label{Deltabar_cw}
\eeq
and we denote $y_{cr}=1-x_{cr}=1/2$ at the lower end of the conformal
window.

In the conformal window, the anomalous dimension at the IRFP in the conformal
window, the exact expression for $\gamma_{\bar\psi\psi,IR}=\gamma_{M,IR}$, is  
\beqs
\gamma_{M,IR} &=& \frac{3C_A}{2T_fN_f} - 1 = \frac{N_{f,u}}{N_f}-1 \cr\cr
              &=&\frac{1}{x}-1 \ .
\label{gamma_Ma}
\eeqs
This can be seen, for example, by solving for $\gamma_M$ at the IR
zero of the NSVZ beta function in Eq. (\ref{beta_nsvz}), which is thus
$\gamma_{M,IR}$. (Another
derivation makes use of the $R$ charges of the $\Phi$ and $\tilde \Phi$
chiral superfields, as discussed in \cite{dexss}.) This anomalous dimension
$\gamma_{M,IR}$ can be expressed in terms of $y$ as follows:
\beq
\gamma_{M,IR} = \frac{\Delta_f}{N_f} = \frac{\Delta_f}{N_{f,u}-\Delta_f}
= \frac{y}{1-y} = \sum_{j=1}^\infty y^j \ . 
\label{gamma_Mb}
\eeq
Thus, the coefficient $\kappa_{M,j}$ in Eq. (\ref{gamma_M_dpdef}) has the
value 
\beq
\kappa_{M,j} = \kappa_{\bar\psi\psi,j} = \frac{1}{(N_{f,u})^j} \ .
\label{kappaj}
\eeq
The finite sum (\ref{gamma_M_dpdef}) was evaluated in our previous work
\cite{dexss}, yielding 
\beq
\gamma_{M,IR,\Delta_f^p} = \gamma_{\bar\psi\psi,IR,\Delta_f^p} =
y\bigg ( \frac{y^p-1}{y-1} \bigg ) \ . 
\label{gamma_dpform}
\eeq
Note that the numerator of the expression on the right-hand side of
Eq. (\ref{gamma_dpform}) contains a factor $(y-1)$ which
cancels the denominator in Eq. (\ref{gamma_dpform}), so that the resulting
expression is a polynomial, as is clear from its definition
(\ref{gamma_M_dpdef}) or from Eq. (\ref{gamma_Mb}).

In \cite{pgb} we showed that, for a given $N_f$ in the conformal
window, $\gamma_{M,IR,\Delta_f^p}$ approaches the exact result in
Eqs. (\ref{gamma_Ma}) and (\ref{gamma_Mb}) exponentially rapidly
(see Eqs. (2.37)-(2.41) in \cite{pgb}).  We recall this result, since it
is relevant here.  As in \cite{pgb}, we define the fractional difference
\beq
\epsilon_p \equiv
\frac{\gamma_{M,IR}-\gamma_{M,IR,\Delta_f^p}}{\gamma_{M,IR}} \ .
\label{epsilon_susy}
\eeq
Using $\gamma_{M,IR}$ from Eq. (\ref{gamma_Ma}) or (\ref{gamma_Mb}) and
$\gamma_{M,IR,\Delta_f^p}$ from Eq. (\ref{gamma_dpform}), this is
\beq
\epsilon_p = y^p \ . 
\label{epsform}
\eeq
Since $y^p = e^{-p\ln(1/y)}$ and $0 < y \le 1/2$ in the conformal
window, this fractional difference evidently approaches zero
exponentially rapidly as a function of the truncation order, $p$. This
is true for any value of $y$ in the conformal window,
and, as a special case, it is true in the limit $y \to y_{cr}=1/2$.

% =================================================================

\section{Anomalous Dimension Conditions In Conformal Window}
\label{gamma_crit_section}

From analyses of the Schwinger-Dyson equation for the fermion 
propagator, of operator product expansions, and other
arguments \cite{alm,cohen_georgi,kaplan_etal,zwicky}, it has been
suggested that the upper bound
\beq
\gamma_{\bar \psi \psi, IR} \le 1
\label{gamma_le_1}
\eeq
applies for an IRFP in the conformal window.  Since
$\gamma_{\bar\psi\psi,IR}$ increases as one decreases $N_f$ throughout
the conformal window, it follows that the lower end of this conformal
regime occurs when the inequality (\ref{gamma_le_1}) is saturated,
i.e., when the following condition holds: 
\beq
\gamma_{\bar\psi\psi,IR}=1 \ .
\label{eqlin}
\eeq
That is, Eq. (\ref{eqlin}) determines the value of $N_{f,cr}$
demarcating the lower end of the conformal window. We denote
Eq. (\ref{eqlin}) as the linear $\gamma$ critical condition, denoted
as L$\gamma$CC.  Note that this condition is in accord with the
exactly known value of $\gamma_{\bar\psi\psi,IR}=\gamma_{M,IR}$ in the
present ${\cal N}=1$ supersymmetric gauge theory, as is clear from the
exact result (\ref{gamma_Ma}).

The quadratic condition 
\beq
\gamma_{\bar \psi \psi,IR}(2-\gamma_{\bar \psi \psi ,IR})=1
\label{eqquad}
\eeq
was discussed as a critical condition for fermion condensation, and its
connection with the condition (\ref{eqlin}) was noted in \cite{alm}
(see also \cite{as_susy}).  We denote Eq. (\ref{eqquad}) as the
quadratic $\gamma$ critical condition, Q$\gamma$CC. As is obvious from
the fact that Eq. (\ref{eqquad}) can be rewritten as
$(\gamma_{\bar\psi\psi,IR}-1)^2=0$, it has a double root at
$\gamma_{\bar \psi \psi,IR}=1$. Hence, an exact solution of the
quadratic equation (\ref{eqquad}) yields the same result as the linear
condition (\ref{eqlin}). However, when applied in the context of
series expansions such as Eq. (\ref{gamma_IR_Deltaseries}), as
calculated to finite order, the results differ from those obtained
with the linear condition (\ref{eqlin}). This difference arises
because the quadratic condition (\ref{eqquad}) generates higher-order
terms in powers of the scheme-independent expansion variable, and
leads to different coefficients of lower-order terms \cite{khl,jwlee}.
In a nonsupersymmetric gauge theory with $N_f$ fermions transforming
according to a single representation of the gauge group, the use of
the quadratic condition (\ref{eqquad}) was found \cite{khl,jwlee} to
(i) show better convergence as a function of increasing order of
truncation of the series (\ref{gamma_IR_Deltaseries}) than the linear
condition (\ref{eqlin}) and (ii) predict a larger value of $N_{f,cr}$
than the linear $\gamma$CC. This work in \cite{khl,jwlee} used the
general results \cite{gsi,dexl} for
$\gamma_{\bar\psi\psi,IR,\Delta_f^p}$ to the highest order that we had
calculated them, namely $p=4$.

As noted in the introduction, an interesting question that we will
investigate here is whether the quadratic $\gamma$CC also converges
more rapidly than the linear $\gamma$CC in the above-mentioned ${\cal
  N}=1$ supersymmetric gauge theory. An additional question that we
will also investigate concerns whether the values of $N_{f,cr}$
obtained from the L$\gamma$CC and Q$\gamma$CC approach the exact value
$N_{f,cr}=3N_c$ from above or below. Equivalently, we will determine
whether the corresponding values of $x_{cr}$ approach the exact value
$x_{cr}=1/2$ from above or from below.  It is worthwhile to mention that
a rigorous upper bound on $\gamma_{\bar \psi \psi,IR}$ in a conformal
field theory is that \cite{mack,gir,nakayama}
\beq
\gamma_{\bar \psi \psi, IR}   \le 2 \ .
\label{cft_bound}
\eeq
This is evidently less restrictive than the bound (\ref{gamma_le_1}). 

% ===================================================================

\section{Calculational Results}
\label{results_section}

The linear $\gamma$CC equation $\gamma_{\bar\psi\psi,IR}-1=0$ with
$\gamma_{\bar\psi\psi,IR}$ calculated to order $O(\Delta_f^p)$
inclusive is Eq. (\ref{eqlin}).  Substituting
Eq. (\ref{gamma_dpform}), this becomes
\beqs
y \bigg ( \frac{y^p-1}{y-1} \bigg ) -1 = 0 \ , 
\label{eqlin_dpa}
\eeqs
or equivalently,
\beq
{\rm L}\gamma{\rm CC}_p: \quad \Big (\sum_{j=1}^p y^j \Big ) -1 = 0 \ .
\label{eqlin_dp}
\eeq
This L$\gamma$CC$_p$ condition is a polynomial equation of degree $p$ in
the variable $y$, or equivalently in the variable $x=1-y$. We denote
the (physical) solution of the L$\gamma$CC equation (\ref{eqlin_dp}),
expressed in terms of the variable $x$, as $x_{cr,L,p}$. 
For the $1 \le p \le 3$, we give the analytic solutions below,
with floating-point values displayed to the indicated
number of significant figures:
\beq
x_{cr,L,1}=0
\label{xlin_d1}
\eeq
\beq
x_{cr,L,2} = \frac{3-\sqrt{5}}{2} = 0.38197
\label{xlin_d2}
\eeq
and
\beqs
x_{cr,L,3} &=& \frac{1}{3}\bigg [ -(17+3\sqrt{33} \, )^{1/3}
  +2(17+3\sqrt{33} \, )^{-1/3} + 4 \bigg ] \cr\cr
&=& 0.456311
\label{xlin_d3}
\eeqs
Although the L$\gamma$CC$_p$ condition (\ref{eqlin_dp}) has $p$ formal
solutions, in each case, there is no ambiguity concerning which of
these is the physical solution. For example, for $p=2$, the other
solution, namely $x=(1/2)(3+\sqrt{5}) = 2.618$ is outside
the conformal-window range, $1/2 \le x < 1$; for $p=3$, the other two
solutions form an unphysical complex-conjugate pair, and so forth for
higher $p$.

The quadratic $\gamma$CC condition (\ref{eqquad}) with
$\gamma_{\bar\psi\psi,IR}$ calculated to $O(\Delta_f^p)$ is 
$(\gamma_{\bar\psi\psi,IR,\Delta_f^p}-1)^2 = 0$. If one takes the square root
of this equation to begin with, one simply recovers the linear $\gamma$CC
equation.  If, instead, one evaluates terms at $O(\Delta_f^p)$ resulting
from the quadratic expression, then one obtains the equation 
\beq
S-1 = 0 \ ,
\label{eqquad_dp}
\eeq
where the sum $S$ has the form 
\beq
S = \sum_{j=1}^p \lambda_j \Delta_f^j \ ,
\label{s}
\eeq
where the coefficients $\lambda_j$ will be discussed shortly. 
Given an input for $\gamma_{\bar\psi\psi,IR}$
calculated to $O(\Delta_f^p)$, the quadratic $\gamma$CC generates
terms up to $O(\Delta_f^{2p})$; however, for self-consistency, one
performs the corresponding truncation of terms to $O(\Delta_f^p)$,
since this is the accuracy of the input expressions for
$\gamma_{M,IR,\Delta_f^p}$. For the coefficients
$\lambda_j$ we calculate that 
\beq
\lambda_1 = 2\kappa_1
\label{lam1}
\eeq
\beq
\lambda_2 = 2\kappa_2 - \kappa_1^2
\label{lam2}
\eeq
\beq
\lambda_3 = 2(\kappa_3 -\kappa_1\kappa_2)
\label{lam3}
\eeq
\beq
\lambda_4 = 2\kappa_4-2\kappa_1 \kappa_3 -\kappa_2^2
\label{lam4}
\eeq
\beq
\lambda_5 = 2(\kappa_5 -\kappa_1 \kappa_4 -\kappa_2\kappa_3) \ , 
\label{lam5}
\eeq
and so forth for higher $j$. In general, we find that $\lambda_j$
contains a term $2\kappa_j$ and then (a) if $j$ is odd, a sum of terms of the
form $-2\kappa_r\kappa_{j-r}$ where $1 \le r \le (j-1)/2$, and (b) if $j$ is
even, a sum of terms of the form $-2\kappa_r \kappa_{j-r}$ with
$1 \le r \le (j/2)-1$, together with a term $-\kappa_{j/2}^2$. Substituting
the expression $\kappa_j = 1/(N_{f,u})^j$ from Eq. (\ref{kappaj}), we find 
\beq
\lambda_j = \frac{3-j}{(N_{f,u})^j}
\label{lamj}
\eeq
and hence
\beq
S = \sum_{j=1}^p (3-j) y^j \ .
\label{sq}
\eeq
Calculating this sum in closed form, we obtain 
\begin{widetext}
\beq
S = \frac{y}{(1-y)^2}\, \bigg [2-3y+(p-2)y^p +(3-p)y^{p+1} \bigg ] \ .
\label{sqform}
\eeq
The numerator of the expression on the right-hand side of
Eq. (\ref{sqform}) contains a factor of $(1-y)^2$ which cancels the
factor in the denominator, so that the result is a polynomial in $y$,
as is obvious from its definition, Eq. (\ref{s}), or from
Eq. (\ref{sq}).  The resultant quadratic $\gamma$CC condition,
evaluated to $O(\Delta_f^p)$, is
\beq
    {\rm Q}\gamma{CC}_p: \quad
    S-1 = \frac{1}{(1-y)^2} \, \bigg [ -1+4y-4y^2+(p-2)y^{p+1}
  +(3-p)y^{p+2} \bigg ] = 0 \ .
\label{sq_minus_1}
\eeq
\end{widetext}
Since $S$ is a polynomial in $y$, it follows that $S-1$ is also, and
hence the expression in square brackets in Eq. (\ref{sq_minus_1})
contains a factor of $(1-y)^2$, which cancels with the $(1-y)^2$ in
the denominator. We denote the (physical) solution of the Q$\gamma$CC
eqution (\ref{sq_minus_1}), expressed in terms of the variable $x$, as
$x_{cr,Q,p}$.  As is clear from Eq. (\ref{sq}), if $p \ne 3$, then the
Q$\gamma$CC$_p$ condition is a polynomial equation of degree $p$ in
the variable $y$, or equivalently in the variable $x$, while if $p=3$,
then the coefficient of the highest-power term vanishes, so the
resultant equation is of degree 2 in $y$.  Indeed, with this
cancellation, the Q$\gamma$CC$_3$ equation is identical to the
Q$\gamma$CC$_2$ equation.  As was the case with the L$\gamma$CC$_p$
condition, although for $p\ge 2$, there are several solutions, there
is no ambiguity concerning which is the physical solution; for
example, for $p=2$, the other solution is $x=2+\sqrt{2}=3.414$, which
is outside the conformal-window range of $x$.  The analytic solutions
to the lowest cases are
\beq
x_{cr,Q,1}=\frac{1}{2}
\label{xcrq1}
\eeq
and
\beq
x_{cr,Q,2}=2-\sqrt{2} =0.58579 \ . 
\label{xcrq2}
\eeq
It happens that the lowest-order result $x_{cr,Q,1}$ is exact, but this
is not generic; for $p \ge 2$, the Q$\gamma$CC$_p$ equation yields a
value of $x_{cr,Q,p} > 1/2$. 

In Table \ref{xlin_dp_table}, we list the results of the calculations
with the linear $\gamma$CC with the input value of
$\gamma_{\bar\psi\psi,IR,\Delta_f^p}$ for 
$1 \le p \le 10$, yielding the L$\gamma$CC$_p$ condition. Table
\ref{xlin_dp_table} includes:

\begin{enumerate}

\item

  the value of $x_{cr,L,p}$,

\item

  the ratio of $x_{cr,L,p}$ to the exact value
$x_{cr}=1/2$, denoted as
\beq
r_{cr,L,p} \equiv \frac{x_{cr,L,p}}{x_{cr}} = 2x_{cr,L,p} \ ,
\label{rlin_dp}
\eeq

\item

  the fractional difference with respect to the exact value,

\beq  
{\rm Diff}_{cr,L,p} \equiv 1 - \frac{x_{cr,L,p}}{x_{cr}} = 1-2x_{cr,L,p} \ , 
\label{diff_Lp}
\eeq

\item

  the fractional difference with respect to the next lower-order value, 

\beq
{\rm Diff}_{cr,L,p,p-1} \equiv 1 - \frac{x_{cr,L,p-1}}{x_{cr,L,p}} \ .
\label{diff_Lppminus1}
\eeq

\end{enumerate}

\begin{table}
\caption{\footnotesize{
    In this table, the columns list 
    (1) the value $p$ specifying the order $O(\Delta_f^p)$ to which the
    linear (L) criticality condition L$\gamma$CC is
    evaluated, yielding the L$\gamma$CC$_p$ condition (\ref{eqlin_dp});
    (2) the value of $N_{f,cr}/N_c$
    calculated from this L$\gamma$CC$_p$ condition, denoted $x_{cr,L,p}$; 
    (3)  the ratio $r_{cr,L,p}$ in Eq. (\ref{rlin_dp});
        (4) the fractional difference with respect to the exact value,
    ${\rm Diff}_{cr,L,p}$ in Eq. (\ref{diff_Lp}); and 
    (5) the fractional difference with respect to the next lower-order value, 
    ${\rm Diff}_{cr,L,p,p-1}$ in Eq. (\ref{diff_Lppminus1}). 
    The abbreviation NA means ``not applicable'', and the notation
    $0.91197$e-2 means $0.91197 \times 10^{-2}$.}}
\begin{center}
\begin{tabular}{|c|c|c|c|c|} \hline\hline
  $p$ & $x_{cr,L,p}$ & $r_{cr,L,p}$ & ${\rm Diff}_{cr,L,p}$ &
  ${\rm Diff}_{cr,L,p,p-1}$ 
\\ \hline
1 & 0        & 0       &  1           & NA         \\
2 & 0.38197  & 0.76393 &  0.23607     & 1          \\
3 & 0.45631  & 0.91262 &  0.087378    & 0.16293    \\
4 & 0.48121  & 0.96242 &  0.037580    & 0.051742   \\
5 & 0.49134  & 0.98268 &  0.017321    & 0.020616   \\
6 & 0.49586  & 0.99172 &  0.82765e-2  & 0.91197e-2 \\
7 & 0.49798  & 0.99597 &  0.40342e-2  & 0.42596e-2 \\
8 & 0.49901  & 0.99801 &  1.98836e-3  & 2.0498e-3  \\
9 & 0.49951  & 0.99901 &  0.98624e-3  & 1.0031e-3  \\
10& 0.49975  & 0.99951 &  0.49092e-3  & 0.49556e-3  \\
\hline\hline
\end{tabular}
\end{center}
\label{xlin_dp_table}
\end{table}

In Table \ref{xquad_dp_table}, we list the results of the calculations
with the quadratic $\gamma$CC with the input value of
$\gamma_{\bar\psi\psi,IR,\Delta_f^p}$ for 
$1 \le p \le 10$, yielding the Q$\gamma$CC$_p$ condition. This table includes:

\begin{enumerate}

\item

  the value of $x_{cr,Q,p}$,

\item

  the ratio of $x_{cr,Q,p}$ to the exact value
$x_{cr}=1/2$, denoted as
\beq
r_{cr,Q,p} \equiv \frac{x_{cr,Q,p}}{x_{cr}} = 2x_{cr,L,p} \ ,
\label{rquad_dp}
\eeq

\item

  the fractional difference with respect to the exact value,

\beq  
{\rm Diff}_{cr,Q,p} \equiv 1 - \frac{x_{cr,Q,p}}{x_{cr}} = 1-2x_{cr,Q,p} \ , 
\label{diff_Qp}
\eeq

\item

  the fractional difference with respect to the next lower-order value, 

\beq
{\rm Diff}_{cr,Q,p,p-1} \equiv 1 - \frac{x_{cr,Q,p-1}}{x_{cr,Q,p}} \ .
\label{diff_Qppminus1}
\eeq

\end{enumerate}

\begin{table}
  \caption{\footnotesize{
    In this table, the columns list 
    (1) the value $p$ specifying the order $O(\Delta_f^p)$ to which the
    quadratic criticality condition Q$\gamma$CC is
    evaluated, yielding the Q$\gamma$CC$_p$ condition (\ref{sq_minus_1});
    (2) the value of $N_{f,cr}/N_c$
    calculated from this Q$\gamma$CC$_p$ condition, denoted $x_{cr,Q,p}$; 
    (3)  the ratio $r_{cr,Q,p}$ in Eq. (\ref{rlin_dp});
        (4) the fractional difference with respect to the exact value,
    ${\rm Diff}_{cr,Q,p}$ in Eq. (\ref{diff_Lp}); and 
    (5) the fractional difference with respect to the next lower-order value, 
    ${\rm Diff}_{cr,Q,p,p-1}$ in Eq. (\ref{diff_Lppminus1}). 
    Other notation is as in Table \ref{xlin_dp_table}.}}
\begin{center}
\begin{tabular}{|c|c|c|c|c|} \hline\hline
  $p$ & $x_{cr,Q,p}$ & $r_{cr,Q,p}$ & ${\rm Diff}_{cr,Q,p}$ &
  ${\rm Diff}_{cr,Q,p,p-1}$ 
\\ \hline
1 & 0.5      & 1       &  0           & NA            \\
2 & 0.58579  & 1.17157 & $-0.17157$   & 0.14644       \\
3 & 0.58579  & 1.17157 & $-0.17157$   & 0             \\
4 & 0.57421  & 1.14843 & $-0.14843$   & $-0.020155$   \\
5 & 0.56145  & 1.12289 & $-0.12289$   & $-0.022738$   \\
6 & 0.54982  & 1.09964 & $-0.09964$   & $-0.021147$   \\
7 & 0.53985  & 1.07969 & $-0.079692$  & $-0.018475$   \\
8 & 0.53153  & 1.06305 & $-0.063050$  & $-0.0156545$  \\
9 & 0.52470  & 1.04940 & $-0.049404$  & $-0.013004$  \\
10& 0.51918  & 1.03836 & $-0.038361$  & $-0.010635$  \\
\hline\hline
\end{tabular}
\end{center}
\label{xquad_dp_table}
\end{table}

We see that in this theory, (i) for a given order $O(\Delta_f^p)$ with
$p \ge 3$, the linear $\gamma$CC yields a value of $x_{cr,L,p}$ that
is closer to the exact value $x_{cr}=1/2$ than the value $x_{cr,Q,p}$
obtained from the quadratic $\gamma$CC, so that the linear $\gamma$CC
yields an estimate of $x_{cr}$ that approaches the exact value more
rapidly than the the estimate from the quadratic $\gamma$CC.  This is
our main result. Furthermore, while the linear $\gamma$CC yields a
value of $x_{cr,L,p}$ that approaches the exact value from below,
the quadratic $\gamma$CC at order $p \ge 2$ yields a value of $x_{cr,Q,p}$
that approaches the exact value from above.  These findings are evident
in Tables \ref{xlin_dp_table} and \ref{xquad_dp_table}. We have checked that
these properties also hold at higher truncation order beyond the
highest order, $p=10$, shown in these tables.

Contrasting these results with those in the corresponding
non-supersymmetric gauge theory, one must first recall that the value
of $N_{f,cr}$ (depending on the gauge group $G$ and the fermion
representation ${\cal R}$) is not known exactly, so that one cannot
make a precise comparison with it.  However, one can, at least,
determine the fractional changes in the values of the solutions for
$x_{cr,L,p}$ and $x_{cr,Q,p}$ as functions of the order $O(\Delta_f^p)$
to which one has calculated $\gamma_{\bar\psi\psi,IR}$.  At an IRFP in
a non-supersymmetric gauge theory with fermions in one representation,
the maximum order to which the scheme-independent calculations have
been performed is $p=4$, with results given in our
Refs. \cite{gsi,dexl}.  It was found in \cite{khl,jwlee} (and
confirmed in \cite{dexm2}), using these results for
$\gamma_{\bar\psi\psi,\Delta_f^p}$ from \cite{gsi,dexl}, that the
quadratic $\gamma$CC converges more rapidly than the linear
$\gamma$CC.  Thus, for $p \ge 3$, the relative accuracies and
convergence rates of the linear versus the quadratic $\gamma$CC that
we find for this ${\cal N}=1$ supersymmetric theory are opposite to
the behavior that was found in the non-supersymmetric
theory. Moreover, in the non-supersymmetric gauge theory, the linear
and quadratic $\gamma$CC conditions yield estimates of $N_{f,cr}$ that
increase as a function of the truncation order, $p$
\cite{khl,jwlee}. This is also true for the values of $N_{f,cr}$ and
thus $x_{cr,L,p}$ obtained from the L$\gamma$CC$_p$ equation in the
supersymmetric gauge theory studied here, i.e., $x_{cr,L,p}$
approaches the exact value $x_{cr}=1/2$ from below. In contrast, in
this supersymmetric theory, for $p \ge 2$ the value of $x_{cr,Q,p}$
calculated from the Q$\gamma$CC$_p$ equation approaches the exact
value of $x_{cr}$ from above.

% =======================================================================

\section{Conclusions}
\label{conclusions_section}

In conclusion, in this paper we have performed a comparison of the
linear and quadratic critical conditions $\gamma_{\bar\psi\psi,IR}=1$
and $\gamma_{\bar\psi\psi,IR}(2-\gamma_{\bar\psi\psi.IR})=1$, where
$\gamma_{\bar\psi\psi,IR}$ is the anomalous dimension of the fermion
bilinear $\bar\psi\psi$ at an infrared fixed point in the conformal
window in an ${\cal N}=1$ supersymmetric gauge theory with $N_f$ pairs
of chiral superfields $\Phi_i$ and $\tilde \Phi_i$ transforming
according to the ${\cal R}$ and $\bar {\cal R}$ representations of the
gauge group $G$, respectively.  This theory has the appeal that both
$\gamma_{\bar\psi\psi,IR}$ and the value $N_{f,cr}$ at the lower
boundary of the conformal window are known exactly.  We find that, as
a function of the order $O(\Delta_f^p)$ to which one uses the
truncated calculation of $\gamma_{\bar\psi\psi,IR}$ as input, for $p
\ge 3$, the linear critical condition yields an estimate of
$x_{cr}=N_{f,cr}/N_{f,u}$ that is more accurate than the quadratic
critical condition.  This behavior is opposite to what was found for
non-supersymmetric gauge theories.  It should be emphasized that the
use of both the linear and quadratic critical conditions with
finite-order inputs for $\gamma_{\bar\psi\psi,IR,\Delta_f^p}$ are
approximate perturbative methods.  Thus, differences between
predictions for the lower end of the conformal window obtained with
these methods provide one measure of the importance of higher-order
terms in the inputs, $\gamma_{\bar\psi\psi,IR,\Delta_f^{p'}}$ with $p'
> p$.  Studies that elucidate the properties of IR-conformal gauge
theories and, in particular, the location of the lower boundary of the
conformal window in these theories, are of continuing interest, both
for basic quantum field theory and for possible phenomenological
applications.  The comparative analysis reported herein provides some
further insight into predictions from different critical conditions
for the lower boundary of the conformal window.

% =====================================================================

\begin{acknowledgments}

  This research of R.S. was supported in part by the U.S. NSF Grant
  NSF-PHY-22-10533.

\end{acknowledgments}

% =======================================================================
% =======================================================================


\begin{thebibliography}{99}
\bibitem{group_invariants}
The group invariants are defined as follows.
Let $T^a_R$ denote the generators of the Lie algebra of $G$ in the
representation $R$, where $a$ is a group index, and let $d_R$ denote the
dimension of $R$.
Then $C_2(R)$ and $T_R$ are given by $T^a_RT^a_R = C_2(R) I$, where here $I$ is
the $d_R \times d_R$ identity matrix, and
${\rm Tr}_R(T^a_R T^b_R) = T(R)\delta^{ab}$.  For a field $f$
transforming according to a representation $R$, we use the equivalent
compact notation $T_f \equiv T(R)$ and $C_f \equiv C_2(R)$. We also use the
notation $C_A \equiv C_2(Adj) \equiv C_2(G)$.
\end{thebibliography}
\end{document}